\documentclass[12pt]{article}
\usepackage[a4paper,margin=0.9in]{geometry}
\usepackage{amsmath}
\usepackage{url}
\usepackage{natbib}
\usepackage{color}
\usepackage{graphicx}
\usepackage{calc}
\usepackage{colordvi}

\input{tcilatex}

\begin{document}

\begin{abstract}
Sequential measurements of non-commuting observables produce order effects
that are well-known in quantum physics. But their conceptual basis, a
significant measurement interaction, is relevant for far more general
situations. We argue that non-commutativity is ubiquitous in psychology
where almost every interaction with a mental system changes that system in
an uncontrollable fashion. Psychological order effects for sequential
measurements are therefore to be expected as a rule. In this paper we focus
on the theoretical basis of such effects. We classify several families of
order effects theoretically, relate them to psychological observations, and
predict effects yet to be discovered empirically. We assess the complexity,
related to the predictive power, of particular (Hilbert space) models of
order effects and discuss possible limitations of such models.
\end{abstract}

%%%%%%%%%%%%%%%%%%%%%%%%%%%%%%%%%%%%%%%%%%%%%%%%%%%%%%%%%%%%%%%%%%%%%%%%%%%%%%%%
%
%%%%%%%%%%%%%%%%%%%%%%%%%%%%%%%%%%%%%%%%%%%%%%%%%%%%%%%%%%%%%%%%%%%%%%%%%%%%%%%%

\begin{titlepage}

\vspace*{\fill}

\bfseries\Large
Order Effects in Sequential Measurements \\[0.5ex] of Non-Commuting Psychological Observables \\[10mm]
\mdseries\large
H.~Atmanspacher$^{1,2}$ and H.~R\"omer$^3$\\[1mm]
\slshape\normalsize
1 - Institute for Frontier Areas of Psychology and Mental Health, Freiburg,
Germany\\
2 - Collegium Helveticum, Zurich, Switzerland \\
3 - Physics Department, University of Freiburg, Germany
\\[10mm]
\mdseries\normalsize\today

\vspace*{\fill}

\tableofcontents

\end{titlepage}

%%%%%%%%%%%%%%%%%%%%%%%%%%%%%%%%%%%%%%%%%%%%%%%%%%%%%%%%%%%%%%%%%%%%%%%%%%%%%%%%
%
%%%%%%%%%%%%%%%%%%%%%%%%%%%%%%%%%%%%%%%%%%%%%%%%%%%%%%%%%%%%%%%%%%%%%%%%%%%%%%%%

\section{Introduction}

\subsection{Measurement and Non-Commutativity}

From a conceptual point of view, measurements are operations including
interactions between a measuring system $M$, e.g.~a measuring device, and a
measured system $S$. Typical objects of a measurement are properties
(sometimes called observables) $A$, $B$, ..., of system $S$ in a certain
state $\rho$. Measurement results for observables $A$ and $B$ are typically
quantified by real-valued numbers $a$, $b$, ...

A physical standard example for this terminology is a material measurement
apparatus $M$ recording particular properties of a material system $S$ in
state $\rho$ (with some probability $p_i$). In basic physics, observables
occur in canonically conjugate pairs, such as position and momentum,
exhibiting particular well-defined invariance principles and conservation
laws.

In psychology we are concerned with a mental system $S$, usually of an
individual subject, being in a state $\rho$. In contrast to physics, there
are no canonically conjugate pairs of observables. For instance, a survey
questionnaire as a measuring device $M$ can contain any questions one may
want to choose as observables, to be measured in terms of answers as
measurement results.

If the measurement interaction between $M$ and $S$ is weak, then the action
of $M$ leaves no significant effect on $S$. In case of strong interactions,
this is no longer the case, and effects of measurements due to $M$ have a
non-negligible influence even on a state $\rho$ of $S$. (This brief
discussion disregards that measurement affects also the state of the
measuring device. This, of course, is intended and mandatory because
otherwise there would be no recorded measurement result.)

In physics, the distinction of weak and strong interactions is one major
criterion for delineating classical systems from quantum systems. In
addition, quantum measurements -- in contrast to classical measurements --
are not simply registrations of pre-existing facts, but they also
``establish'' the fact which is registered. The key feature of this
classical-quantum distinction is mathematically codified in terms of the
commutativity or non-commutativity, respectively, of observables. Referring
to measurements as actions of an observable on a state, $A(\rho)$ or $B(\rho)
$, these actions either leave the state as it was before measurement, or
they change it.

As a consequence, the successive action of two observables on a state can be
commutative, $A(B(\rho)) = B(A(\rho))$, or non-commutative, $A(B(\rho)) \neq
B(A(\rho))$. In physics, the measured behavior of a system is called \textit{%
classical} in the former case, whereas the latter case refers to \textit{%
quantum} behavior. From this perspective, quantum theory can be viewed as a
general systems theory of both commuting and non-commuting observables, and
classical physics is restricted to the special case of commuting observables
alone (cf.~Primas 1990).

Whenever the sequence of successive measurement interactions between a
measuring device and a measured system makes a difference for the final
result, measurements are non-commutative. Many complex systems must be
expected to exhibit non-commutative properties in this sense. In psychology,
where virtually every interaction of a ``measuring'' device with a
``measured'' mental state changes that state uncontrollably and where mental
states are often literally established by measurements, it is highly
plausible to argue that non-commutativity should be the ubiquitous rule.

\subsection{Non-Commutative Phenomena in Psychology}

Traditionally, psychology has not considered its observations as owing to
effects of non-commu\-ta\-tive observables throughout its history. But there
are quite a number of psychological phenomena showing clear features of such
an approach. Some of them, which have been worked out in recent years, are
decision (or judgment) processes, semantic associations, bistable
perception, learning, and order effects in questionnaires.\footnote{%
For some more details and references see the review by Atmanspacher (2011),
which highlights this line of thinking in Sec.~4.7. Ideas to understand
consciousness in terms of actual quantum processes in the brain, also
reviewed in the same article, are entirely at variance with the approach
applied here.}

It should be noted at this point that the models developed in the mentioned
areas are not all \textit{explicitly} based on non-commuting observables.
They feature various related concepts such as incompatibility,
complementarity, entanglement, partial Boolean logic, dispersive states,
quantum probability, uncertainty relations etc. However, these concepts bear
close formal relationships with the property of non-commutativity and its
ramifications (cf.~Atmanspacher \textit{et al.} 2002, 2006, Filk and R\"omer
2011; more technically: R\'edei 1996, R\'edei and Summers 2007).

A particular one among the psychological phenomena mentioned above is most
illustratively addressed in terms of non-commuting operations: the
phenomenon of order effects. These effects do so obviously refer to
non-commutativity because they are ruled by it almost literally. For two
questions (actions of observables) $A$ and $B$, their sequence makes a
difference with respect to the answer (state $\rho$ of the system) after the
second question,
\begin{equation}
AB\rho \neq BA\rho \, ,  \notag
\end{equation}
where $A$ and $B$ are operations whose actions on the state $\rho$ of the
system change $\rho$. Related to this, there are no proper joint probability
distributions of $A$ and $B$ in the sense of standard statistical theory,
but only order-dependent probabilities to be defined in Sec.~2.

Psychologists know such effects well, they are also called sequence effects
or context effects. An (incomplete) list of textbooks and reviews on the
topic, showing its fairly long history, is Sudman and Bradburn (1974),
Schumann and Presser (1981), Schwarz and Sudman (1992), Hogarth and Einhorn
(1992), Sudman, Bradburn and Schwarz (1996), Tourangeau, Rips, and Rasinski
(2000). However, the proposed models for describing order effects have
always been framed classically (e.g., Markov processes, Bayesian updating,
etc.). Models inspired by quantum theory have been successfully applied only
recently.

The first publication with corresponding indications that we know of has
been due to Aerts and Aerts (1995) and essentially derived from a
non-Boolean logic of propositions. With the growing success of and attention
on non-classical modeling over the last decade, more work has been done
along those lines. Most notable with respect to our topic in the present
paper are applications to sequential decision making by Pothos and Busemeyer
(2009) Busemeyer \textit{et al.} (2011), Trueblood and Busemeyer (2011), and
Wang and Busemeyer (2011), whose work basically utilize quantum
probabilities in Hilbert space representations of mental states.

Our own formal framework for addressing non-commuting properties is somewhat
more abstract, because the full-fledged formalism of (Hilbert space) quantum
mechanics is generally too restrictive for modeling non-commutative
situations.\footnote{%
In Sec.~4.3 we point out how non-commutative behavior not describable in
Hilbert space quantum mechanics may arise in psychological systems.} There
are scenarios in which we cannot even require such basic concepts as a
metric (linear or nonlinear) representation space for states, the
formulation of dynamical laws, or a definition of probabilities. Of course,
some of these may become necessary if specific examples are to be treated
and empirical results are to be described. We have shown elsewhere
(Atmanspacher \textit{et al.} 2002, 2006, Filk and R\"omer 2011) how
conventional quantum mechanics can be stepwise recovered if enough relevant
contexts are filled into the general framework.

It is to be expected that models of psychological phenomena should be
located somewhere in between the most general level and the very specific
level of conventional quantum physics. For instance, psychology will
arguably not have a universal quantity, such as the Planck constant,
specifying the degree of non-commutativity in general. On the other hand,
contact with controlled experiments will definitely make it necessary to
have well-defined probabilities.

Hence, we begin this article with a brief account of probabilities for the
results of sequential measurements of non-commuting observables in Section
2. Section 3 uses this framework for a theoretically framed classification
of different types of order effects, some of which have been discussed in
the psychological literature; others are predicted here for the first time.
Section 4 addresses Hilbert space models of such order effects. It is shown
that the complexity of such models, if they fit empirically obtained
probabilities, is lower than the number of parameters to be fitted. In other
words, the models have considerable predictive power.\footnote{%
The complexity of a model is related to the number of its parameters. The
model is parsimonious if this number is smaller than the number of variables
to be fitted. The predictive power of a model increases with its parsimony.}
Finally we give an example for the limits of Hilbert space models of order
effects. Section 5 summarizes the results of this article.

%%%%%%%%%%%%%%%%%%%%%%%%%%%%%%%%%%%%%%%%%%%%%%%%%%%%%%%%%%%%%%%%%%%%%%%

\section{Probabilities for Sequential Measurements}

Let a system be in state $\rho$ before its observables $A$ and $B$ are
measured. Then the \textit{individual measurement probability} (IMP) $%
w_{\rho}(a_j)$ or $w_{\rho}(b_i)$ is the probability of measuring $a_j$ of $A
$ or $b_i$ of $B$ in state $\rho$, without any prior measurement influence.

Now consider subsequent measurements of first $A$ and then $B$ in the
initial state $\rho$ and denote the probability of measuring the value $b_i$
of $B$ after having measured value $a_j$ of $A$ as the \textit{sequential
measurement probability} (SMP) $w_{\rho}(b_i \leftarrow a_j)$. If $A$ and $B$
do not commute, $w_{\rho}(b_i \leftarrow a_j)$ will be different from $%
w_{\rho}(a_j \leftarrow b_i)$ (cf.~Franco 2009). For the sake of simplicity,
we assume $a_j$ and $b_i$ to be discrete.

The SMP is the fundamental quantity in the discussion to follow. It is not
simply the conditional probability of measuring $b_i$ given $a_j$ but refers
to subsequent measurements of both $A$ and $B$ that actually have been
carried out and, thus, may (and generally will) change the state $\rho$ of
the system. This will be discussed in some more detail in Sec.~4.1.

Since $w$ is a probability (in the sense of Kolmogorov), the sum of $%
w_{\rho}(b_i \leftarrow a_j) $ over all measurements of $A$ and $B$ must be
normalized:
\begin{equation}
\sum_{i,j} w_{\rho}(b_i \leftarrow a_j) = 1
\end{equation}
Moreover, the sum of $w_{\rho}(b_i \leftarrow a_j) $ over all measured
values $b_i$ of $B$ is given by
\begin{equation}
\sum_i w_{\rho}(b_i \leftarrow a_j) = w_{\rho}(a_j) \, .
\end{equation}

The conditional probability
\begin{equation}
w_{\rho}(b_i|a_j) := \frac{1}{w_{\rho}(a_j)} \, w_{\rho}(b_i \leftarrow a_j)
\end{equation}
in principle contains information about both ``ordinary'' statistical
dependence and effects of the non-commutativity of observables $A$ and $B$.
In the non-commutative case, Bayes' rule is violated:
\begin{equation}
\frac{w_{\rho}(b_i|a_j)}{w_{\rho}(a_j|b_i)} \neq \frac{w_{\rho}(b_i)} {%
w_{\rho}(a_j) }
\end{equation}

Concentrating on effects due to non-commuting observables alone and
disregarding ordinary statistical dependence, we need to consider ``pooled''
probabilities $w_{A,\rho}(b_i)$ , summing over all possible values $a_j$ of $%
A$, rather than conditional probabilities as in Eq.~(3):
\begin{equation}
w_{A,\rho}(b_i) := \sum_j w_{\rho}(b_i \leftarrow a_j) \, .
\end{equation}
If $w_{A,\rho}(b_i) \neq w_{\rho}(b_i)$, we have an unambiguous criterion
for non-commuting observables. (Of course, $a_j$ and $b_i$ in relations (1)
to (5) must be exchanged if the sequence of measuring $A$ and $B$ is
swapped.)

It follows that the expectation value $\langle B \rangle_{\rho}$ for
measuring $B$ alone (unconditioned) differs from the expectation value $%
\langle B \rangle_{a_j, \rho}$ for measuring $B$ after $A$:\footnote{%
Throughout this paper, we use the notation $\langle . \rangle$ for
expectation values, also known as (disregarding subtle details) mean values
of a distribution.}
\begin{equation}
\langle B \rangle_{a_j, \rho} := \sum_i b_i \, w_{\rho}(b_i|a_j) \neq
\langle B \rangle_{\rho} = \sum_i b_i \, w_{\rho}(b_i) \, ,
\end{equation}
which in general also holds for any function $f(B)$ and its expectation
values for both commuting or non-commuting observables. Again, concentrating
on effects of non-commutativity we have to use pooled probabilities and get:
\begin{equation}
\langle B \rangle_{A, \rho} := \sum_i b_i \, w_{A,\rho}(b_i) \neq \langle B
\rangle_{\rho} = \sum_i b_i \, w_{\rho}(b_i) \, .
\end{equation}

Likewise, the expectation values of the variance operators $(\Delta A)^2$
and $(\Delta B)^2$, where the operators $\Delta A$ and $\Delta B$ are the
deviations from the expectation values of $A$ and $B$, differ as well. In
the case of both commuting or non-commuting observables, we have in general:
\begin{equation}
\langle (\Delta B)^2 \rangle_{a_j, \rho} = \langle (B-\langle B
\rangle_{a_j, \rho})^2\rangle_{a_j, \rho} \ \neq \ \langle (B-\langle B
\rangle_{\rho})^2\rangle_{\rho} = \langle (\Delta B)^2 \rangle_{\rho}
\end{equation}
Concentrating on effects of non-commutivity alone, with pooled
probabilities, this turns into:
\begin{equation}
\langle (\Delta B)^2 \rangle_{A, \rho} = \langle (B-\langle B \rangle_{A,
\rho})^2\rangle_{A, \rho} \ \neq \ \langle (B-\langle B
\rangle_{\rho})^2\rangle_{\rho} = \langle (\Delta B)^2 \rangle_{\rho}
\end{equation}

Furthermore we can define a joint sequential expectation value as:
\begin{equation}
\langle B \leftarrow A \rangle_{\rho} = \sum_{ij} a_j \, b_i \, w_{\rho}(b_i
\leftarrow a_j)
\end{equation}

In general (for instance in quantum mechanics), the expectation value
\begin{equation}
\langle \Delta A \, \Delta B \rangle_{\rho} = \langle (A- \langle A
\rangle_{\rho})(B- \langle B \rangle_{\rho})\rangle_{\rho}
\end{equation}
may not even be real-valued because the product operator $\Delta A \, \Delta
B$ is not Hermitian if $A$ and $B$ do not commute. This is at variance with
classical (commuting) observables as used in standard statistical theory,
where the product of real-valued observables is always real-valued.

Now we define
\begin{eqnarray}
\delta_{\rho} (b_i) &=& \sum_j w_{\rho}(b_i \leftarrow a_j) - w_{\rho}(b_i)
\\
\delta_{\rho} (a_j) &=& \sum_i w_{\rho}(a_j \leftarrow b_i) - w_{\rho}(a_j)
\end{eqnarray}
as the differences between sums of SMPs and IMPs for measuring $A$ and $B$.
These differences provide a convenient way to assess the effect of
sequential measurements of $A$ and $B$ as compared to individual
measurements of $A$ or $B$ alone. It may be mentioned that the transition
probabilities for sequential measurements as introduced here represent a
special case of Khrennikov's (2009) \textit{contextual} probabilities.

%%%%%%%%%%%%%%%%%%%%%%%%%%%%%%%%%%%%%%%%%%%%%%%%%%%%%%%%%%%%%%%%%%%%%%%%%%%%%%%%

\section{Types of Order Effects}

\subsection{Theoretical Classification of Observed Order Effects}

With the definitions in Eqs.~(12,13), we can relate our probabilistic
framework to consistency and contrast effects first reported by Schumann and
Presser (1981) and to additive and subtractive effects found by Moore
(2002). All four types have recently been investigated by Wang and Busemeyer
(2011), see also Trueblood and Busemeyer (2011) and Busemeyer \textit{et al.}
(2011).

\textit{Additive effects} (both SMPs are larger than the corresponding
IMPs):
\begin{equation}
\delta_{\rho} (a_j) > 0 \quad \mathrm{and} \quad \delta_{\rho} (b_i) > 0
\end{equation}
\textit{Subtractive effects} (both SMPs are smaller than the corresponding
IMPs):
\begin{equation}
\delta_{\rho} (a_j) < 0 \quad \mathrm{and} \quad \delta_{\rho} (b_i) < 0
\end{equation}
\textit{Contrast effects} (the difference of the two SMPs is larger than the
difference of the corresponding IMPs):
\begin{equation}
\left|\sum_i w_{\rho}(a_j \leftarrow b_i) - \sum_j w_{\rho}(b_i \leftarrow
a_j)\right| \ > \ \left|w_{\rho}(a_j) - w_{\rho}(b_i)\right|
\end{equation}
\textit{Consistency effects} (the difference of the two SMPs is smaller than
the difference of the corresponding IMPs):
\begin{equation}
\left|\sum_i w_{\rho}(a_j \leftarrow b_i) - \sum_j w_{\rho}(b_i \leftarrow
a_j)\right| \ < \ \left|w_{\rho}(a_j) - w_{\rho}(b_i) \right|
\end{equation}

Contrast and consistency effects apply in particular to situations where $%
\delta_{\rho} (a_j)$ and $\delta_{\rho} (b_i)$ have different signs, $%
\delta_{\rho} (a_j) \delta_{\rho} (b_i) < 0$. In this case, additive or
subtractive effects are not candidates anyway since there the product is
always positive. Of course, all inequalitites turn into equalities if $A$
and $B$ commute.

For a comprehensive discussion of pertinent examples of these four classes
of order effects see Wang and Busemeyer (2011). They developed an approach
based on quantum probabilities (cf.~Sec.~4) and were able to describe
empirical results of earlier surveys with excellent accuracy, far exceeding
that of alternative classical models.

\subsection{Further Options for Future Application}

In addition to the four kinds of order effects mentioned above, there are
other interesting possibilities. For instance, non-commuting observables
imply a general asymmetry of SMPs
\begin{equation}
w_{\rho}(a_j \leftarrow b_i) - w_{\rho}(b_i \leftarrow a_j) \neq 0 \, ,
\end{equation}
which serves as a measure for the degree to which $A$ and $B$ do not
commute, i.e.~$AB\neq BA$. If (and only if) an addition of observables is
well-defined, this entails a non-vanishing commutator, $AB-BA\neq 0$.

Considering the variances according to Eq.~(8), with non-pooled
probabilities, two further effects can be distinguished. If
\begin{equation}
\langle (\Delta B)^2 \rangle_{a_j, \rho} < \langle (\Delta B)^2
\rangle_{\rho} \, ,
\end{equation}
this may be characterized as a \textit{contraction effect}: the variance of
measuring $B$ decreases if $A$ is measured first. Contraction effects
indicate the degree to which $A$ and $B$ are interdependent
(i.e.~compatible). The alternative case is:
\begin{equation}
\langle (\Delta B)^2 \rangle_{a_j, \rho} > \langle (\Delta B)^2
\rangle_{\rho} \, ,
\end{equation}
meaning that the variance of measuring $B$ increases if $A$ is measured
first, characterizing a \textit{distraction effect}. Distraction effects
indicate the degree to which $A$ and $B$ are incompatible
(i.e.~non-commutative).

For the specifically non-commutative case, we must again replace $\langle .
\rangle_{a_j,\rho}$ by $\langle . \rangle_{A,\rho}$ as in Eq.~(9), so that
contraction is characterized by
\begin{equation}
\langle (\Delta B)^2 \rangle_{A, \rho} < \langle (\Delta B)^2 \rangle_{\rho}
\, ,
\end{equation}
and distraction is characterized by
\begin{equation}
\langle (\Delta B)^2 \rangle_{A, \rho} > \langle (\Delta B)^2 \rangle_{\rho}
\, .
\end{equation}

Distraction effects are typical in quantum mechanics, e.g.~for non-commuting
observables such as position $Q$ and momentum $P$ of a quantum system. A
measurement designed to fix $Q$ as precisely as possible entails an
increased variance of $P$. In quantum mechanics this is expressed by
Heisenberg-type uncertainty relations, where the product of the two
variances is bounded from below by their commutator.

Although numerous order effects in questionnaires, surveys or polls have
been found with respect to shifted expectation values, studies of
uncertainty relations with respect to variances are more difficult than
studies of mean shifts and have not been carried out so far. If such
uncertainty relations were empirically discovered, the lower bound of the
product of the variances of the distribution of sequential responses might
provide an estimate for the degree to which the questions considered do not
commute.

A further useful measure for testing order effects is due to correlations
between measurements. Defining, as in quantum theory, ${\frac{1}{2}} \langle
\Delta A \, \Delta B + \Delta B \, \Delta A \rangle_{\rho}$ for correlations
between individual measurements and ${\frac{1}{2}}(\langle \Delta A
\leftarrow \Delta B\rangle_{\rho} + \langle \Delta B \leftarrow \Delta A
\rangle_{\rho})$ for correlations between sequential measurements, two kinds
of order effects are possible: \newline
\textit{Correlation enhancement}:
\begin{equation}
\langle \Delta A \, \Delta B + \Delta B \, \Delta A \rangle_{\rho} < \langle
\Delta A \leftarrow \Delta B\rangle_{\rho} + \langle \Delta B \leftarrow
\Delta A \rangle_{\rho}
\end{equation}
\textit{Correlation attenuation}:
\begin{equation}
\langle \Delta A \, \Delta B + \Delta B \, \Delta A \rangle_{\rho} > \langle
\Delta A \leftarrow \Delta B\rangle_{\rho} + \langle \Delta B \leftarrow
\Delta A \rangle_{\rho}
\end{equation}
The symmetrized product observable ${\frac{1}{2}} \langle \Delta A \, \Delta
B + \Delta B \, \Delta A \rangle_{\rho}$ is well-defined in quantum theory,
but its operationalization is not straightforward because it is not feasible
by sequential measurements.

\section{Order Effects in Hilbert Space Representations}

\subsection{Sequential Measurements in Hilbert Space Quantum Mechanics}

In ordinary Hilbert space quantum mechanics (von Neumann 1932), both pure
and mixed states can be represented by density matrices, i.e.~positive
normalized self-adjoint operators $\rho$ with $\rho^+ = \rho$, $\rho > 0$, $%
tr \rho =1$. \textit{Pure} states are states of \textit{individual systems},
typically represented by state vectors $\psi$ in a Hilbert space. Every
superposition of pure states represents another pure state. Pure states
encode maximal information about the system. \textit{Mixed} states are
states of \textit{statistical ensembles} of pure states with different
probabilities, also called statistical states. Pure states can be
represented by density operators that are idempotent, $\rho = \rho^2$.

In ordinary Hilbert space representations, the IMPs $w_{\rho}(a_j)$ and $%
w_{\rho}(b_i)$ are defined as
\begin{eqnarray}
w_{\rho}(a_j) &=& tr (P_{a_j} \rho) \ = \ tr (P_{a_j} \rho P_{a_j}) \, , \\
w_{\rho}(b_i) &=& tr (P_{b_i} \rho) \ = \ tr (P_{b_i} \rho P_{b_i}) \, ,
\end{eqnarray}
where $P_{a_j}$ and $P_{b_i}$ are the projection operators onto eigenstates
of $A$ and $B$ with corresponding eigenvalues $a_j$ and $b_i$. An eigenstate
of an observable $A$ with eigenvalue $a_j$ is a state in which a measurement
of $A$ yields $a_j$ with probability 1.

The density matrix $\rho_{a_j}$, i.e.~the state of the system after the
measurement result of $a_j$ has been obtained, is:
\begin{equation}
\rho_{a_j} = \frac{P_{a_j} \rho P_{a_j}}{ tr (P_{a_j} \rho P_{a_j}) }
\end{equation}

The (conditional) probability for measuring $b_i$ after measuring $a_j$ is:
\begin{equation}
w_{\rho} (b_i | a_j) \ = \ tr P_{b_i} \rho_{a_j} \ = \ \frac{tr (P_{b_i}
P_{a_j} \rho P_{a_j} P_{b_i} )} {tr ( P_{a_j} \rho P_{a_j} )} \ = \
w_{\rho_{a_j}} (b_i)
\end{equation}
As a consequence, the (sequential) probability of measuring first $a_j$ and
then $b_i$ is given by:
\begin{equation}
w_{\rho} (b_i \leftarrow a_j) \ = \ w_{\rho} (a_j) w_{\rho} (b_i | a_j) \ =
\ tr (P_{b_i} P_{a_j} \rho P_{a_j} P_{b_i} )
\end{equation}
Likewise, we have
\begin{equation}
w_{\rho} (c_k \leftarrow b_i \leftarrow a_j) \ = \ tr (P_{c_k} P_{b_i}
P_{a_j} \rho P_{a_j} P_{b_i} P_{c_k})
\end{equation}
for three successive measurements, and analogously for more.

\subsection{Complexity of Hilbert Space Models}

In this subsection, we investigate how the number of model parameters needed
to fit experimental data depends on the dimension of the Hilbert space used
to represent the states. The studies by Moore (2002) were based on binary
alternatives, hence Wang and Busemeyer (2011) used a two-dimensional Hilbert
space for fitting them. Furthermore, the number of parameters depends on
whether pure states or mixed states are admitted. Wang and Busemeyer (2011)
based their analysis on pure states. (If the states of different individuals
are not assumed to be identical, it might be reasonable to employ mixed
states.)

Let us first consider the situation in a two-dimensional Hilbert space.
Choosing normalized eigenstates of $A$ as a basis, pure states are
characterized by two real-valued parameters, because pure states are
represented by complex two-dimensional vectors $\psi \neq 0$ (modulo a
multiplicative complex number). By contrast, mixed states have three
real-valued parameters, because $\rho$ is a self-adjoint $2\times 2$ matrix
with $tr \rho = 1$.

The relative position of a self-adjoint $2\times 2$ matrix $B$ with respect
to $A$ as a reference (corresponding to the transformation from the
eigenspaces of $A$ to the eigenspaces of $B$) is determined by two
real-valued parameters. Altogether, this results in four real-valued
parameters for pure states, and five for mixed states.

What needs to be fitted are the measured probabilities $w_{\rho}(a_1)$, $%
w_{\rho}(b_1)$, $w_{\rho}(a_1 \leftarrow b_1)$, $w_{\rho}(a_1 \leftarrow b_2)
$, $w_{\rho}(b_1 \leftarrow a_1)$, and $w_{\rho}(b_1 \leftarrow a_2)$ --
compare Eqs.~(1) and (2). Hence, six empirically obtained numbers have to
fitted by four (respectively five) model parameters, which documents that
two-dimensional Hilbert space models provide a non-trivial compact
description of the empirical data to be fitted.

In an $n$-dimensional Hilbert space with $n> 2$ possible measurement results
for each observable, the situation is analogous. For a pure (mixed,
respectively) state we need $2n-2$ ($n^2-1$, respectively) real-valued
parameters plus $n^2-n$ parameters for the relative positions. Together this
provides $n^2+n-2$ parameters for pure states, and $2n^2-n-1$ for mixed
states.

The number of probabilities to be fitted for $i,j = 1, ..., n$ are $2n-2$
for the IMPs and $2n^2 -2n$ for the SMPs (compare the 2-dimensional case),
hence $2n^2-2$ in total. So the difference between parameters to be fitted
and model parameters is $n(n-1)$ for pure states, and $n-1$ for mixed
states. Hence, the model becomes more parsimonious with increasing $n$. As a
consequence, its relative complexity (with respect to the number of
parameters to be fitted) decreases with increasing $n$, and it decreases
particularly fast for pure states.\footnote{%
The estimated numbers may be further reducible if additional knowledge is
available about the state and the relative positions of $A$ and $B$. The
complexity corresponding to the numbers given thus refers to the most
general and least parsimonious situation within the class of Hilbert space
models.} This indicates an increasing predictive power of the model with
increasing $n$.

The predictive power increases dramatically if more than two observables are
considered. For instance, three observables on an $n$-dimensional Hilbert
space give rise to $2n^2-2$ (pure states) or $3n^2-2n-1$ (mixed states)
model parameters. The number of empirical probabilities to be fitted is $%
3(n-1) + 6n(n-1) + 6n^2(n-1)$ in this case. For $n=2$ this amounts to 39
probabilities to be fitted by 6 (7) model parameters for pure (mixed) states.

For empirical applications, this result suggests that scaled responses in
questionnaires should be favored over binary responses to exploit the
increasing parsimony of higher-dimensional Hilbert space models. In this
case, however, one should keep in mind that the total number of parameters
increases with Hilbert space dimension and may lead to overly high
variances. In practice this can be balanced by a large number of
observations.

In a finite-dimensional Hilbert space model, the state of the system
determines the expectation values and variances of all observables,
including the case of sequential measurements. The state $\rho$ is given by
a finite number of real-valued parameters fixed by the Hilbert space
dimension. By contrast, the state of a classical system is described by a
multivariate distribution function which can require many parameters. Their
number may be reducible by regarding only restricted families of
distributions, but for order effects changes of distributions must also be
taken into account, in turn increasing the number of model parameters.
Models based on stochastic processes, e.g.~hidden Markov models, are typical
examples of such situations.

In contrast, Hilbert space models have a tendency to be less complex (more
parsimonious) because only the state and the relative positions of
observables enter into the number of model parameters. This should become
visible for concrete models of specific situtations. Comparing Hilbert space
modeling for decision making (Busemeyer \textit{et al.} 2011) with classical
probability models (Bayesian model selection), Shiffrin and Busemeyer (2011)
found indeed that the complexity is less for Hilbert space models. It should
be noted, however, that they could not explore the entire parameter space
and their results have not yet been generalized beyond the one experimental
situation they looked at.

\subsection{Joint Measurements}

In the standard formalism of quantum theory by von Neumann (1932), two
observables $A$ and $B$ can be jointly measured (i.e.~attributed definite
values) without restriction only if they commute.\footnote{%
More precisely, this holds for ``ideal'' (von Neumann) measurements ``of the
first kind'' (Pauli). For realistic laboratory experiments with measurements
``of the second kind'' positive-operator-valued measures of ``unsharp''
observables have been suggested by Davies, Holevo, and Ludwig in the 1970s.
See Uffink (1994) for a critical discussion of ``unsharp'' joint
measurements of non-commuting observables, and for further references.} This
commutativity corresponds to the fact that the algebra of measurements is
Boolean and, thus, expresses decisions of \textit{binary} alternatives. In a
Hilbert space representation, measurements are projections onto subspaces of
the Hilbert space, and observables are usually defined as projection-valued
self-adjoint operators.

Joint measurements of observables $A$ and $B$ that need not commute can be
described in the following way: Let $P_{a_j}$ and $P_{b_i}$ be the
propositions that a measurement of $A$ yields $a_j$ and a measurement of $B$
yields $b_i$. Then,
\begin{equation}
P_{a_j \wedge b_i} = \lim_{n\rightarrow \infty} (P_{a_j} P_{b_i})^n
\end{equation}
can be shown to be the projection onto the (symmetric) intersection of the
associated eigenspaces of $A$ and $B$ and corresponds to the conjunction
(i.e.~the fusion of two questions into one in a questionnaire)
\begin{equation}
P_{a_j \wedge b_i} = P_{b_i \wedge a_j} \, .
\end{equation}
Note that the limit in Eq.~(31) does not coincide with the product $(P_{a_j}
P_{b_i})$ or $(P_{b_i} P_{a_j})$ in the non-commutative case. If no common
eigenstate of $(P_{a_j}$ and $P_{b_i})$ exists, we have $P_{a_j \wedge b_i}
= 0$.

Adding a prior measurement of $a_j$ by $P_{a_j}$, we get
\begin{equation}
P_{a_j} \, P_{a_j \wedge b_i} = P_{a_j \wedge b_i} \, P_{a_j} = P_{a_j
\wedge b_i} = P_{b_i} \, P_{a_j \wedge b_i} = P_{a_j \wedge b_i} \, P_{b_i}
\end{equation}

Because in Hilbert space quantum mechanics probabilities are expressed as
traces of projections on states,
\begin{equation}
w_{\rho}(a_j\wedge b_i \leftarrow a_j) = tr P_{a_j \wedge b_i} P_{a_j} \rho
P_{a_j} = tr P_{a_j \wedge b_i} \rho \, ,
\end{equation}
we obtain the condition:
\begin{equation}
w_{\rho}(a_j\wedge b_i \leftarrow a_j) = w_{\rho} (a_j \wedge b_i) =
w_{\rho}(a_j \leftarrow a_j\wedge b_i)
\end{equation}

Condition (35) is equally obtained if additional measurements are added.
Consider the case of measuring $a_j$ first, then $b_i$, then $a_j \wedge b_i$%
. The concatenation of these measurements in the corresponding SMP gives
\begin{equation}
w_{\rho}(a_j\wedge b_i \leftarrow b_i \leftarrow a_j) = tr P_{a_j \wedge
b_i} P_{b_i} P_{a_j} \rho P_{a_j} P_{b_i} = tr P_{a_j \wedge b_i} \rho =
w_{\rho} (a_j \wedge b_i) \, ,
\end{equation}
yielding again the IMP $w_{\rho} (a_j \wedge b_i)$ for combined
propositions. As a result, Hilbert space quantum mechanics predicts that the
SMP of a sequence of individual and joint measurements always yields the
probability of the joint measurement alone. As a consequence, the
propositions $P_{a_j}$ and $P_{a_j \wedge b_i}$ are always compatible so
that order effects are excluded for $P_{a_j}$ and $P_{a_j \wedge b_i}$ in
the projector calculus of Hilbert space quantum mechanics, and also in
propositional logic.

If $P_{a_j}$ and $P_{a_j \wedge b_i}$ are \textit{empirically} found to be
incompatible, e.g.~as in ``anomalous'' consistency effects or contrastive
effects, this would indicate that standard Hilbert space models with
self-adjoint operators are insufficient to describe the data properly.
Psychological experiments with joint measurements are therefore interesting
candidates for exploring potential limitations of Hilbert space modeling. It
depends on the situation, which kind of extension of the model will be
required or advisable in such cases.

For example, the response to a joint-measurement question $P_{a_j \wedge b_i}
$ does logically determine the response to each individual question $P_{a_j}$
or $P_{b_i}$. On the other hand, asking $P_{a_j}$ or $P_{b_i}$ first may
lead to a response deviating from the one obtained when asking $P_{a_j}
\wedge P_{b_i}$ first.\footnote{%
Using the example of the conjunction fallacy (Tversky and Kahneman 1983), $%
P_{a_j}$ may be the response to ``Linda is a bankteller'', $P_{b_i}$ the
response to ``Linda is a feminist'', and $P_{a_j} \wedge P_{b_i}$ the
response to ``Linda is a feminist backteller''. It is not implausible that
the state of a subject having responded with $P_{a_j} \wedge P_{b_i}$
differs from her state having responded with $P_{a_j}$ to a degree which
creates an ``anomalous'' order effect defying conventional Hilbert space
modeling. It remains to be explored whether or not positive-operator-valued
measures as mentioned in footnote 6 would be capable of resolving such a
case with a proper description.}  This is conceivable if agents behave in
conflict with the rules of Boolean logic. In this case, sequential survey
questions may produce order effects \textit{appearing as if} $P_{a_j}$ and $%
P_{b_i}$ did not commute with their conjunction.

Since in a Hilbert space framework $P_{a_j} \wedge P_{b_i}$ commutes with
both $P_{a_j}$ and $P_{b_i}$, this framework is too narrow for a proper
description of such order effects (with joint measurements). The challenge
of such situations would be to find a non-Hilbert space representation in
which both $P_{a_j}$ and $P_{b_i}$ in fact do not commute with their
conjunction even formally. This would amount to finding a representation
more general than a Hilbert space model for the particular ``anomalous''
behavior observed.

\section{Summary}

In his summary of the volume edited by Schwarz and Sudman (1992), one of the
pioneers of research on sequential measurements in psychology stated
(Bradburn 1992): ``One of the factors that inhibited our progress in
understanding order effects has been the lack of a theoretical structure
within which to investigate the mechanisms by which they might occur.'' The
present paper presents an attempt toward such a theoretical basis.

We consider the non-commutativity of measurement operations as the formal
key to order effects. This is well established in quantum theory where
observables typically do not commute: The sequence in which measurement
operations act upon the state of a system makes a difference for the results
obtained. One reason is that a measurement operation changes the state (even
it is pure) of the measured system such that a subsequent measurement
operation effectively acts on another state. Related to this, a quantum
measurement is not simply the registration of a pre-existing fact, but also
establishes the fact that is registered.

It is highly plausible that this basic idea holds for psychological systems
as well, although states, observables and their dynamcis have nothing to do
with quantum physics. The mathematical feature of non-commuting observables
and its ramifications can be fruitfully applied to model psychological
situations where order effects abound.

We proposed a classification of particular order effects, some of which were
already discovered empirically, and Hilbert space models were successfully
used to analyze them: additive effects, subtractive effects, contrast
effects, and consistency effects. Moreover, we predicted order effects not
observed so far, which rest on uncertainty relations between variances of
distributions of observables rather than on shifts of their expectation
values.

We assessed the complexity of Hilbert space models for pure and mixed
states, and for $n$-dimensional Hilbert spaces corresponding to questions
(observables) with $n$-ary alternatives for response. It turned out that the
predictive power of the model increases with increasing $n$, and that it
increases dramatically with increasing number of observables. We sketched an
argument why the complexity of Hilbert space models should be generally less
than that of alternative classical models.

We suggested experiments with joint measurements as interesting candidates
for exploring potential limitations of Hilbert space models. If agents
behave in a way conflicting with Boolean logical rules, order effects may
result whose description requires non-commutative frameworks more general
than standard Hilbert space models permit.

It is evident that a successful model does not entail the validity of a
mechanism. However, models based on non-commuting observables stress the
specific feature of strong measurement interactions. This important point,
which is highly plausible for observations on mental systems, entails that
non-commutative models are not only descriptively powerful but also hold the
potential for explanatory surplus. In combination with their parsimony over
classical models, non-commutative models offer a very attractive option for
future research.

In contrast to measurements for classical systems, registering values of
observables in both quantum and cognitive systems includes manipulations of
the state of the system: (1) the posed and answered measurement question
brings the system into a particular state in which it has (in general) not
been before the question was asked, and (2) the registration itself entails
a backreaction on the system which changes its state once more, so that it
differs from the state that was actually measured.

This far-reaching issue shows that a quantum theoretically inspired
understanding of cognition is capable of revising plugged-in cliches of
thinking in terms of classically construed concepts. The law of
commutativity in elementary calculations (and the related Boolean
``either--or'' in logic) are special cases with their own significance. But
it would be wrong to believe that their generalization is restricted to
exotic particles and fields in microphysics, with no application for
everyday life phenomena. The opposite is the case.

\section*{Acknowledgments}

We are grateful to three reviewers and, in particular, to Jerome Busemeyer
for constructive criticism and valuable suggestions for improvements.

%%%%%%%%%%%%%%%%%%%%%%%%%%%%%%%%%%%%%%%%%%%%%%%%%%%%%%%%%%%%%%%%%%%%%%%%%%%%%%%%

\section*{References}

\begin{description}
\item Aerts, D., and Aerts S.~(1995): Applications of quantum statistics in
psychological studies of decision processes. \textit{Foundations of Science}
\textbf{1}, 85---97.

\item Atmanspacher, H.~(2011). Quantum approaches to consciousness. \textit{%
Stanford Encyclopedia of Philosophy}, ed.~by E.N.~Zalta, accessible at %
\url{http://plato.stanford.edu/entries/qt-consciousness/}.

\item Atmanspacher, H., R\"omer, H., and Walach, H.~(2002).  Weak quantum
theory: Complementarity and entanglement  in physics and beyond. \textit{%
Foundations of Physics} \textbf{32}, 379--406.

\item Atmanspacher, H., Filk, T., and R\"omer, H.~(2006). Weak quantum
theory: Formal framework and selected applications. In: Adenier, G., \textit{%
et al.} (eds.) \textit{Quantum Theory: Reconsideration  of Foundations - 3},
pp.~34--46, New York: American Institute of Physics.

\item Bradburn, N.N.~(1992). What have we learned? In: Schwarz, N., and
Sudman, S.~(eds.) \textit{Context Effects in Scial and Psychological Research%
}, pp.~315--323, Berlin: Springer.

\item Busemeyer, J.R., Pothos, E., Franco, R., and Trueblood, J.S.~(2011). A
quantum theoretical explanation for probability judgment errors. \textit{%
Psychological Review} \textbf{108}, 193---218.

\item Filk, T., and R\"omer, H.~(2011). Generalized quantum theory: Overview
and latest developments. \textit{Axiomathes} \textbf{21}, 211--220.

\item Franco, R.~(2009). The conjunction fallacy and interference effects.
\textit{Journal of Mathematical Psychology} \textbf{53}, 415--422.

\item Hogarth, R.M., and Einhorn, H.J.~(1992). Order effects in belief
updating. The belief-adjustment model. \textit{Cognitive Psychology} \textbf{%
24}, 1--55.

\item Khrennikov, A.~(2009). \textit{Contextual Approach to Quantum Formalism%
}, Berlin: Springer.

\item Moore, D.W.~(2002). Measuring new types of question-order effects.
\textit{Public Opinion Quarterly} \textbf{66}, 80--91.

\item Pothos, E., and Busemeyer, J.R. (2009). A quantum probability
explanation for violations of ``rational'' decision theory. \textit{%
Proceedings of the Royal Society B} \textbf{276}, 2171--2178.

\item Primas, H.~(1990). Mathematical and philosophical questions in the
theory of open and macroscopic quantum systems. In: A.I.~Miller (ed.)
\textit{Sixty-Two Years of Uncertainty}, pp.~233--257, Berlin: Springer.

\item R\'edei, M.~(1996). Why John von Neumann did not like the Hilbert
space formalism of quantum mechanics  (and what he liked instead). \textit{%
Studies in the History and Philosophy of Modern Physics} \textbf{27},
493--510.

\item R\'edei, M., and Summers, S.J.~(2007). Quantum probability theory.
\textit{Studies in the History and Philosophy of Modern Physics} \textbf{38}%
, 390--417.

\item Schumann, H., and Presser, S.~(1981). \textit{Questions and Answers in
Attitude Surveys: Experiments on Question Form, Wording and Content}, New
York: Academic Press.

\item Schwarz, N., and Sudman S., eds. (1992). \textit{Context Effects in
Social and Psychological Research}, Berlin: Springer.

\item Shiffrin, R.M., and Busemeyer, J.~(2011). Selection of quantum
probability models. Presented at the symposium ``The Potential of Quantum
Probability for Modeling Cognitive Processes'', 33rd Annual Cognitive
Science Conference.

\item Sudman, S., and Bradburn, N.M.~(1974). \textit{Response Effects in
Surveys}, Chicago: Aldine.

\item Sudman, S., Bradburn, N.M., and Schwarz, N.~(1996). \textit{Thinking
about Answers: The Application of Cognitive Processes to Survey Methodology}%
, Jossey-Bass, San Francisco.

\item Trueblood, J., and Busemeyer, J.R.~(2011). A quantum probability
account of order effects in inference. \textit{Cognitive Science}, in press.

\item Tversky, A., and Kahneman, D.~(1983). Extension versus intuitive
reasoning: The conjunction fallacy in probability judgment. \textit{%
Psychological Review} \textbf{90}, 293---315.

\item Uffink, J.~(1994). The joint measurement problem. \textit{%
International Journal of Theoretical Physics} \textbf{33}, 199--212.

\item von Neumann, J.~(1932). \textit{Mathematische Grundlagen der
Quantenmechanik}, Berlin: Springer. English translation: \textit{%
Mathematical Foundations of Quantum Mechanics}, Princeton: University Press
(1955).

\item Wang, Z., and Busemeyer, J.R. (2011).  Explaining and predicting
question order effects using a quantum model. Preprint.
\end{description}

%%%%%%%%%%%%%%%%%%%%%%%%%%%%%%%%%%%%%%%%%%%%%%%%%%%%%%%%%%%%%%%%%%%%%%%%%%%%%%%%

\end{document}